\documentclass[aps,prb,twocolumn,superscriptaddress,showkeys]{revtex4-2}

\usepackage[pdftex]{graphicx}
\usepackage{color,latexsym,inputenc,setspace,multirow,array,tabularx,bm}
\usepackage{amsfonts,amssymb,amsmath,braket}
\usepackage[colorlinks=true]{hyperref}

\newcommand{\bacovo}{BaCo$_{2}$V$_{2}$O$_{8}$}
\newcommand{\co}{Co$^{2+}$}

\begin{document}

\title{Phase transitions and spin dynamics of the quasi-one dimensional\\
Ising-like antiferromagnet \bacovo\ in a longitudinal magnetic field}

\author{Shintaro Takayoshi}
\email[Corresponding author: ]{takayoshi@konan-u.ac.jp}
\affiliation{Department of Physics, Konan University, 658-8501 Kobe, Japan}

\author{Quentin Faure}
\affiliation{Universit\'e Grenoble Alpes, CEA, IRIG/DEPHY/MEM/MDN, 38000 Grenoble, France}
\affiliation{Universit\'e Grenoble Alpes, CNRS, Institut N\'eel, 38042 Grenoble, France}

\author{Virginie Simonet}
\affiliation{Universit\'e Grenoble Alpes, CNRS, Institut N\'eel, 38042 Grenoble, France}

\author{B\'{e}atrice Grenier}
\email[Corresponding author: ]{grenier@ill.fr}
\affiliation{Universit\'e Grenoble Alpes, CEA, IRIG/DEPHY/MEM/MDN, 38000 Grenoble, France}

\author{Sylvain Petit}
\affiliation{Laboratoire L\'eon Brillouin, CEA, CNRS, Universit\'e Paris-Saclay, CEA-Saclay, 91191 Gif-sur-Yvette, France}

\author{Jacques Ollivier}
\affiliation{Institut Laue Langevin, 38000 Grenoble, France}

\author{Pascal Lejay}
\affiliation{Universit\'e Grenoble Alpes, CNRS, Institut N\'eel, 38042 Grenoble, France}

\author{Thierry Giamarchi}
\affiliation{Department of Quantum Matter Physics, University of Geneva, 1211 Geneva, Switzerland}

(Dated: \today)

\begin{abstract}

By combining inelastic neutron scattering and numerical simulations, we study the quasi-one dimensional Ising-like quantum antiferromagnet \bacovo\ in a longitudinal magnetic field applied along the magnetic anisotropy axis, which is also the chain direction. The external field closes the excitation gap due to the magnetic anisotropy, inducing a transition from the N\'eel ordered state to an incommensurate longitudinal spin density wave phase. If the field is increased further, another transition into a transverse antiferromagnetic phase takes place at 9 T due to the competition between longitudinal and transverse correlations. We numerically and experimentally show that the model of XXZ chains connected by a weak interchain interaction well reproduces this transition. We also calculate the dynamical susceptibility and demonstrate that it agrees quantitatively with inelastic neutron scattering measurements. In contrast to the abrupt change of magnetic ordering, the spectra do not change much at the transition at 9 T, and the spin dynamics can be described as a Tomonaga-Luttinger liquid. We also refine the modeling of \bacovo\ by including a four-site periodic term arising from the crystal structure which enables to account for an anomaly of the magnetic susceptibility appearing at 19 T as well as for the anticrossing observed in the inelastic neutron scattering spectra.

\end{abstract}

\maketitle

\section{Introduction}

Low dimensional correlated systems have attracted a wide interest in modern condensed matter physics due to a variety of phenomena arising from both the strong quantum fluctuations and interactions. In particular, one-dimensional (1D) systems are an ideal playground to study many kinds of quantum phases and transitions between them~\cite{giamarchi2004}. Since there are a lot of powerful analytical and numerical methods to study 1D systems, we can deeply understand the properties of strongly correlated quantum systems by using them. For example, effective field theory obtained from bosonization, Bethe Ansatz, numerics by means of matrix product states such as density matrix renormalization group (DMRG)~\cite{white1992}, and time-dependent block decimation (TEBD)~\cite{vidal2003} are quite useful for their investigation. Hence, the experimental realization of 1D systems is important to verify the prediction from theories and clarify their applicability. The compound \bacovo\ is one of such examples. It is well described as a quasi-1D system, specifically antiferromagnetic (AF) XXZ spin chains connected by a weak interchain interaction. Due to the easy-axis anisotropy, the ground state has an AF long range order (N\'eel order) and a spin excitation gap below the N\'eel temperature $T_{\mathrm{N}}=5.6\;\mathrm{K}$. In contrast to the deconfined excitations, called spinons, and their multiple continuum observed in canonical XXZ chains~\cite{descloizeaux,Lake2005}, the interchain interaction in \bacovo\,causes the confinement of spinons, which is detected as a discrete sequence of dispersions in inelastic neutron scattering spectra~\cite{Grenier2015prl,Faure2018}. 

The application of a magnetic field along the crystal ${\bf c}$ axis enriches the emergent phenomena~\cite{Kimura2007,Okunishi2007,Klanjsek2015}. In the case of a pure 1D XXZ chain, when the applied field is increased, the gap closes, inducing a Pokrovsky-Talapov type transition into the Tomonaga-Luttinger liquid (TLL) phase~\cite{Pokrovsky1979,Chitra1997,Giamarchi1999}, which then continues up to the magnetization saturation. This magnetization process can be described by a bosonized field theory. In \bacovo, with the presence of interchain interaction, a three dimensional (3D) long range order is stabilized~\cite{Grenier2015prb,Faure2019}, which corresponds to dominant spin-spin correlations along the XXZ chain. An incommensurate longitudinal order appears in the low field regime and a commensurate transverse one in the high field one. Although most of the features in \bacovo\ can be explained in terms of the XXZ model and a weak interchain coupling, there is still a little complexity stemming from the crystal structure. The spin chains consist in magnetic $\mathrm{Co}^{2+}$ ions that wind to form a screw along the ${\bf c}$ axis. Since the cobalt oxide octrahedra are slightly tilted from the ${\bf c}$ axis, additional perturbation terms appear in the Hamiltonian~\cite{Kimura2007,Kimura2022}. Thus, to make a solid connection between the material and the model to describe it, more detailed research both on the theoretical and experimental sides is necessary. 

In this paper, we study the phase diagram, phase transition, and spin dynamics of \bacovo\ in a longitudinal magnetic field (applied along the anisotropy axis, which is the ${\bf c}$ axis). In particular, we elucidate the mechanism of the transition between the incommensurate longitudinal phase and the commensurate transverse one. Our precise numerical simulations, based on an effective 1D model treating the interchain interaction by a mean field theory, demonstrate that the quantum phase transition is caused by the competition between the longitudinal and transverse correlations within a chain. This conclusion is supported by the very good agreement obtained when comparing the inelastic neutron scattering spectra calculated and measured for different values of the applied magnetic field. In contrast to the abrupt change in the magnetic ordering, the spin dynamics does not vary much across the transition, and can be described as a TLL. These results indicate that the quasi-1D model of XXZ spin chains connected by a weak interchain coupling describes \bacovo\ appropriately. This simple XXZ Hamiltonian, however, cannot capture the anomaly reported in the magnetic susceptibility at a field where the magnetization takes half of the saturated value. To solve this issue, the model was refined by adding four-site periodic perturbations arising from the crystal structure of \bacovo. We find that the refined model not only reproduces this anomaly but also the anticrossing of excitation dispersions observed in inelastic neutron scattering spectra. 

This paper is organized as follows. In Sec.~\ref{sec:Model}, we present the phase diagram of \bacovo\ and detail the quasi-1D model to describe it. We also evaluate the interchain coupling from the numerics with a mean-field approximation. Section~\ref{sec:Result} discusses the evolution of spin dynamics in \bacovo\ under an applied longitudinal magnetic field, on the basis of numerical calculations combined with inelastic neutron scattering measurements. In Sec.~\ref{sec:Refinement}, we refine the model for \bacovo\ to give a better description beyond the simple TLL theory. We summarize our results and discuss future problems in Sec.~\ref{sec:Summary}. Technical details are given in Appendices. 

\section{System and model}
\label{sec:Model}

\subsection{Phase diagram}
\label{sec:ExpPhaseDiag}

\begin{figure}[t]
\centering
\includegraphics[width=8cm]{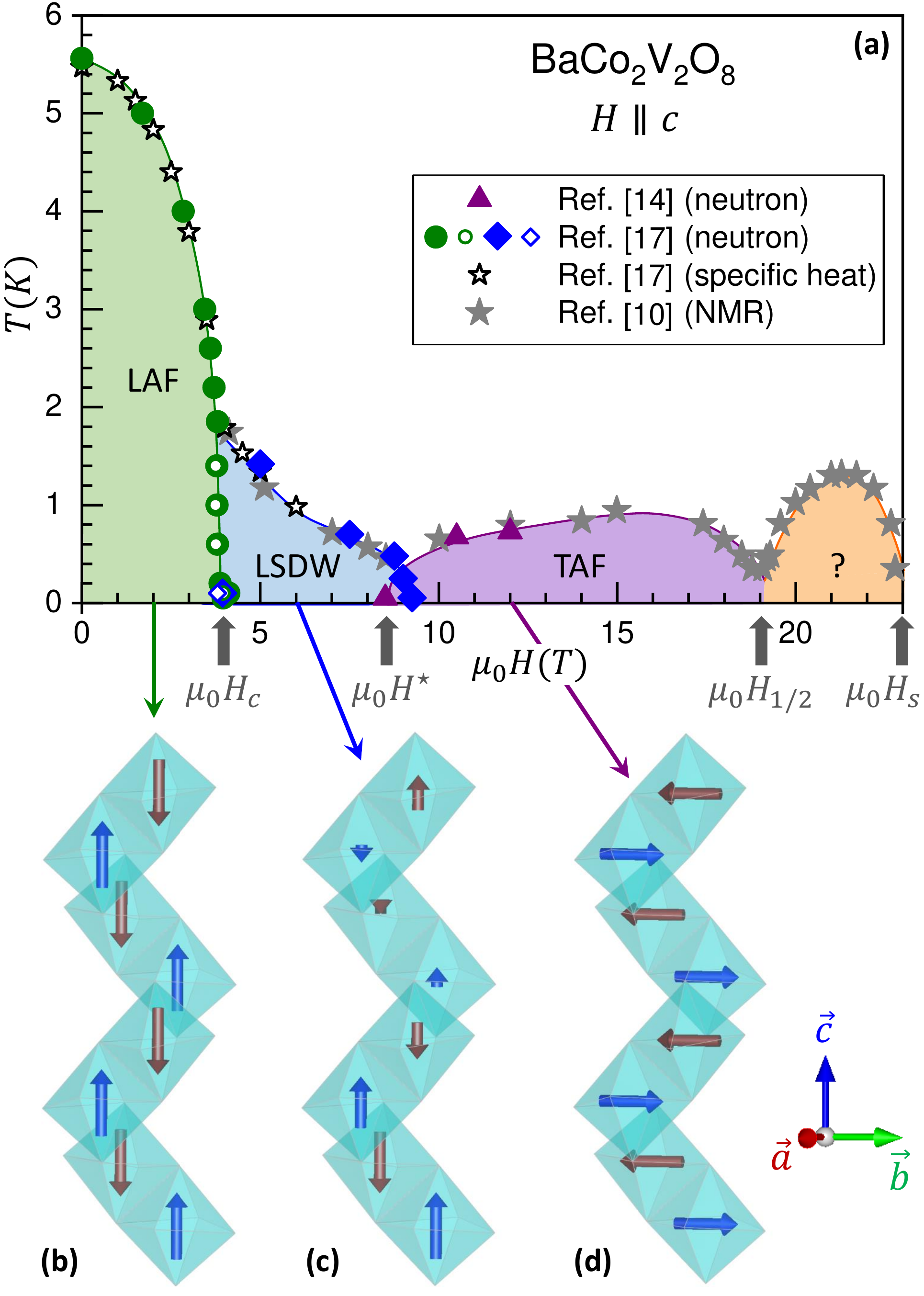}
\caption{(a) Phase diagram of \bacovo\ in the parameter space of the longitudinal magnetic field and temperature determined from neutron diffraction, specific heat, and NMR measurements. The phases LAF, LSDW, and TAF correspond to longitudinal antiferromagnetic, longitudinal spin density wave, and transverse antiferromagnetic orderings, respectively, as determined by neutron diffraction \cite{Canevet2013,Grenier2015prb}, while the phase labeled ``?'' was inferred to be another incommensurate phase from NMR \cite{Klanjsek2015}. (b)-(d) Schematic pictures for the spin arrangement in the magnetic phases LAF, LSDW, and TAF appearing in the field region below the field $H_{1/2}$ corresponding to half of the saturated magnetization. These sketched components should be added to a ferromagnetic component, i.e., uniform magnetization, which increases with the field above $H_{\mathrm{c}}$ [see Fig.~\ref{fig.2}(a), right axis].}
\label{fig.1}
\end{figure}

At low temperature, as the applied longitudinal magnetic field (${\bf H\parallel c}$) is increased, \bacovo\ displays a succession of different magnetic phases through quantum phase transitions. The critical fields of these transitions and the phase diagram [see Fig.~\ref{fig.1}(a)] in the parameter space of the longitudinal magnetic field and temperature have been investigated by magnetization and specific heat~\cite{Kimura2008,Canevet2013}, neutron scattering~\cite{Kimura2008b,Canevet2013,Grenier2015prb,Faure2019}, electron spin resonance (ESR)~\cite{Kimura2007}, and nuclear magnetic resonance (NMR)~\cite{Klanjsek2015}. Neutron diffraction measurements up to 12 T have additionally allowed to identify the spin arrangements in each phase, as sketched in Figs.~\ref{fig.1}(b), \ref{fig.1}(c), and \ref{fig.1}(d)~\cite{Canevet2013,Grenier2015prb}. At $\mu_{0} H_{\mathrm{c}}\simeq 3.9\;\mathrm{T}$, the excitation gap due to the Ising anisotropy closes and the system enters a longitudinal spin density wave (LSDW) phase below 1.8 K, under the influence of the interchain couplings. In this phase, while the ordered magnetic moments are still oriented along ${\bf c}$, the wave vector of the spin density is incommensurate and varies with the amplitude of the field. 
At $\mu_{0}H^{*} \simeq 9\;\mathrm{T}$, another quantum phase transition occurs below $0.7\;\mathrm{K}$ toward a phase characterized by a commensurate AF ordering with magnetic moments lying in the $({\bf a,b})$ plane, i.e., perpendicular to the magnetic field, on top of a uniform component parallel to the field. We call it the transverse antiferromagnetic (TAF) phase.
Furthermore, an anomaly in the magnetic susceptibility is reported around $\mu_{0} H_{1/2}= 19.5\;\mathrm{T}$~\cite{Kimura2007,Klanjsek2015}, which is hypothesized as a transition to another magnetic phase, before the magnetization reaches the saturation field around $\mu_{0} H_{s}=22.7\;\mathrm{T}$. Note that the uniform magnetization along the field is just half of its saturation value at $\mu_{0} H_{1/2}$~\cite{Kimura2007}. This anomaly at $\mu_{0} H_{1/2}=19.5\;\mathrm{T}$ is further discussed in Sec.~\ref{sec:Refinement}. 

Other experimental techniques have contributed to the investigation of 
the spin dynamics, especially inelastic neutron 
scattering~\cite{Grenier2015prl,Grenier2015prlerratum,Faure2019} in the 
ordered longitudinal antiferromagnetic (LAF) and LSDW phases up to 6.8 T and ESR~\cite{Kimura2007} in the LAF and paramagnetic phases up to 25.6 T. The phase diagram, magnetic order, and spin excitations of the sister compound SrCo$_2$V$_2$O$_8$, which presents very similar properties, was also investigated by THz spectroscopy and neutron scattering \cite{Wang2018,Bera2020}.

\subsection{Model}

The properties of \bacovo\ explained in Sec.~\ref{sec:ExpPhaseDiag} can be well described by a quasi-1D model based on the Hamiltonian 
\begin{align}
 \mathcal{H}
   =\mathcal{H}_{\mathrm{intra}}
   +\mathcal{H}_{\mathrm{inter}},
\label{eq:Htot}
\end{align}
where 
\begin{align}
 \mathcal{H}_{\mathrm{intra}}=&
\, J \sum_{n,\mu} 
 [\epsilon
 ( S_{n,\mu}^{x} S_{n+1,\mu}^{x} 
  + S_{n,\mu}^{y} S_{n+1,\mu}^{y})
  + S_{n,\mu}^{z} S_{n+1,\mu}^{z}]
\nonumber\\
  &- g_{zz}\mu_{B}\mu_{0} H\sum_{n,\mu} S_{n,\mu}^{z}
\label{eq:Hintra}
\\
 \mathcal{H}_{\mathrm{inter}}=&
\, J'\sum_{n}\sum_{\braket{\mu,\nu}}
 {\bf S}_{n,\mu} \cdot
 {\bf S}_{n,\nu} 
\end{align}
represent the intrachain and interchain parts of the coupling, respectively. ${\bf S}_{n,\mu}$ is a spin-$1/2$ operator, $n$ is the site index along the chain, $\mu$ and $\nu$ label different chains, $g_{zz}$ is the Land\'e factor, and $\mu_B$ is the Bohr magneton. We employ the values of the intrachain interaction $J= 5.8\; \mathrm{meV}$ and anisotropy parameter $\epsilon=0.53$ already used in previous works~\cite{Faure2018,Faure2019,Faure2021}, and the axes $x$, $y$, and $z$ correspond to the ${\bf a}$, ${\bf b}$, and ${\bf c}$ directions, respectively. 
If there is no interchain interaction, Eq.~\eqref{eq:Hintra} is merely an XXZ model under a longitudinal field, which can be well understood in terms of a bosonized field theory. For details, see Appendix~\ref{sec:bosonization}. With an easy-axis anisotropy, however, there is an excitation gap at zero field. When the field is increased, the gap closes and the transition to the TLL phase occurs. Without any interchain interaction, there is no phase transition until the magnetization is saturated. In the TLL phase, the transverse and longitudinal correlation functions decay as power laws, whose respective exponents depend on the Luttinger parameter $K$.  $K$ is found to increase monotonically with the field from $1/4$ to $1$. Since the leading order of the transverse and longitudinal correlation functions are $r^{-\frac{1}{2K}}$ and $r^{-2K}$, respectively ($r$ is the distance between spins), the longitudinal (transverse) correlations are dominant in the $1/4< K <1/2$ ($1/2< K <1$) regime.

When the interchain interaction is introduced, long-range ordered magnetic phases appear. The LSDW phase is stabilized for the regime of dominant longitudinal correlations while the TAF phase is stabilized for the regime of dominant transverse correlations~\cite{Klanjsek2015}. Therefore we can state that the transition at $\mu_{0} H^{*}\simeq 9\;\mathrm{T}$ in \bacovo\ stems from the competition between the longitudinal and transverse correlations.

\subsection{Phase diagram from numerics}

We further investigate the model given by  Eq.~\eqref{eq:Htot} treating the interchain interaction in terms of a mean field theory. Using this approximation, we derive the following effective 1D XXZ model 
\begin{align}
 \mathcal{H}_{\mathrm{eff}}
   =& J \sum_{n} [\epsilon
 ( S_{n}^{x} S_{n+1}^{x} 
  + S_{n}^{y} S_{n+1}^{y})
  + S_{n}^{z} S_{n+1}^{z}]
\nonumber\\
   &- g_{zz}\mu_{B}\mu_{0} H\sum_{n} S_{n}^{z}
     + J'\sum_{n}
     \braket{{\bf S}_{n}}\cdot
     {\bf S}_{n},
\label{eq:Hamil1Deff}
\end{align}
where $\braket{{\bf S}_{n}}$ is calculated iteratively and determined self-consistently from numerical DMRG calculation. 

\begin{figure}
\centering
\includegraphics[width=8.5cm]{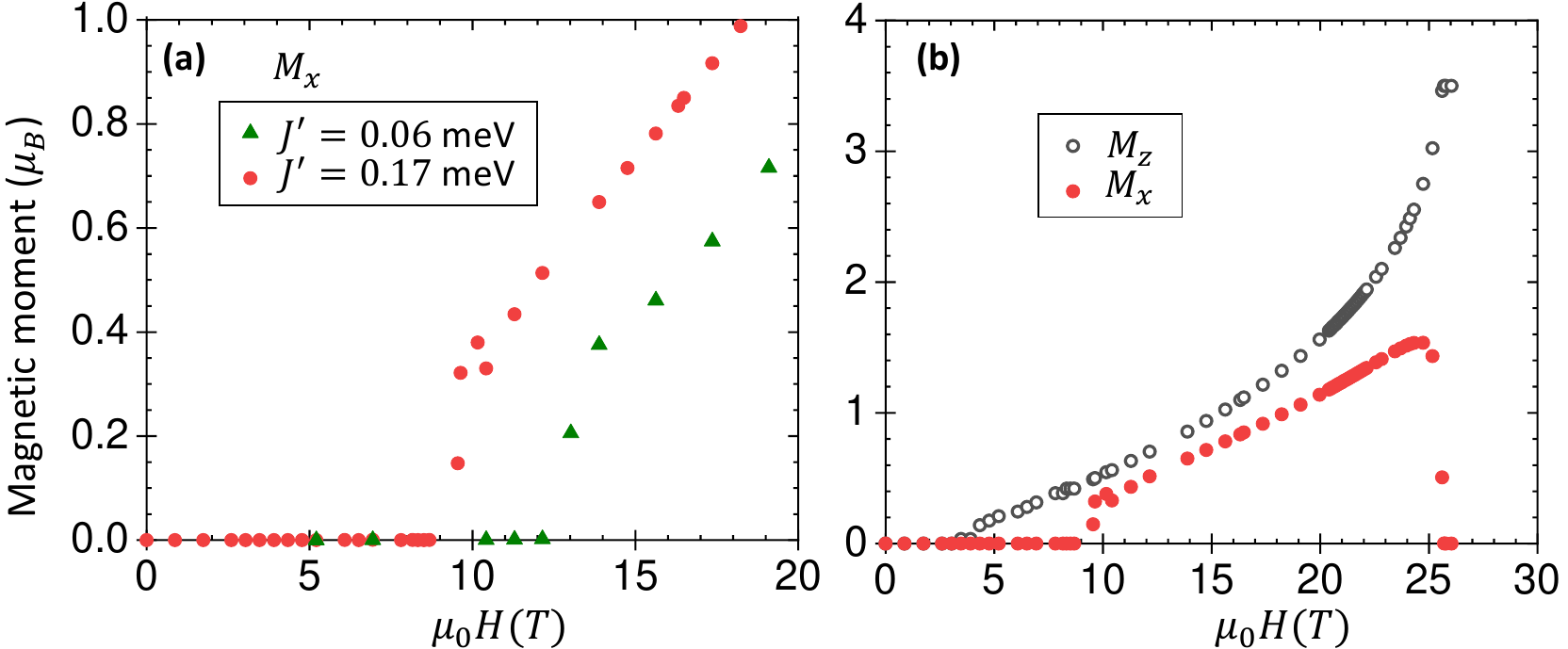}
\caption{(a) Calculated staggered transverse magnetization $M_{x}$ versus the magnetic field for two different interchain interactions $J'$. The spin-flop transition is obtained at 9~T for $J'=0.17$~meV. (b) $M_{x}$ and uniform magnetization $M_{z}$ versus the magnetic field calculated up to saturation for $J'=0.17$~meV.}
\label{fig.4}
\end{figure}

\begin{figure*}
\centering
\includegraphics[width=16cm]{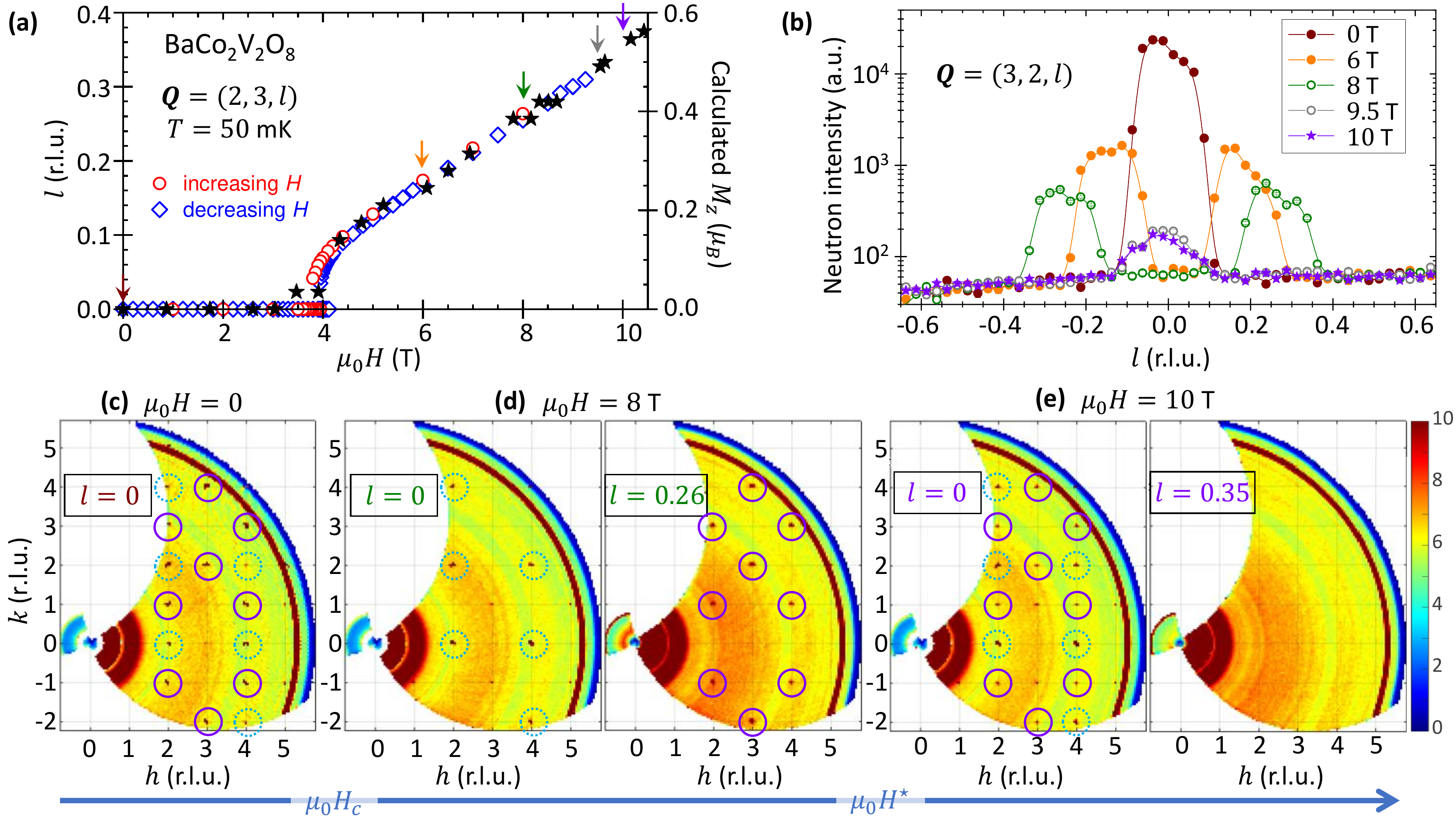}
\caption{(a) Field dependence of the component along ${\bf c^*}$ of the magnetic reflection determined by neutron diffraction (open symbols, left axis) \cite{Canevet2013}. It scales with the calculated field-induced ferromagnetic component, i.e., uniform magnetization, $M_{z}$ (black solid stars, right axis) [see Fig.~\ref{fig.4}(b) and main text]. The arrows show the $l$ positions for the incommensurate modulation $\delta$ in reciprocal lattice units r.l.u., which is probed by neutron diffraction up to 9.3 T, then is extrapolated above for magnetic fields corresponding to the IN5 measurements presented in panel (b). (b) shows $l$-cuts around the $(3, 2, 0)$ position at zero energy, $T=50\; \mathrm{mK}$, and different magnetic fields. The neutron intensity is given in arbitrary units (a.u.). (c)-(e) Intensity maps at constant energy equal to $0.0\pm0.1\;\mathrm{meV}$ in the $(h,k,0)$ and $(h,k,\delta)$ scattering planes, at 50 mK for different magnetic fields: (c) 0 T, (d) 8 T where $\delta = 0.26$ was previously determined in the LSDW phase and (e) 10 T where one would expect incommensurate magnetic Bragg peaks with $\delta = 0.35$ if the system would still be in the LSDW phase. The blue dotted circles highlight the nuclear Bragg peaks and the purple solid circles the magnetic Bragg peaks corresponding either to the $(1,0,0)$ or $(1,0,\delta)$ propagation vector.}
\label{fig.2}
\end{figure*}

As explained above, the pure 1D XXZ spin chain in an external longitudinal field is described as a TLL, thus without long-range order. The interchain interaction is therefore necessary to obtain the spin-flop transition between the two ordered phases: LSDW and TAF. In the mean-field theory, the transverse staggered magnetic order in the TAF phase is stabilized due to an alternating transverse field effectively introduced by the interchain interaction in a self-consistent way. Since the TAF phase becomes more stable as the interchain interaction is larger, the critical field of the spin-flop transition decreases as $J'$ is increased [see Fig.~\ref{fig.4}(a)]. The interchain interaction $J'=0.17\;\mathrm{meV}$ well reproduces the critical field $H^*$ measured experimentally. This value of $J'$ is the same as the one determined in a similar mean-field approach to reproduce the inelastic neutron scattering spectra measured in \bacovo\ under the application of a transverse magnetic field~\cite{Faure2018}. Note that in our previous investigation of the effect of the longitudinal field~\cite{Faure2019}, the value of $J'$ which best reproduces the spectra was $J'=0$. This may be due to the fact that the mean field approach on the interchain interaction is not a good approximation in the vicinity of the LAF-LSDW transition at $\mu_{0}H_{\mathrm{c}}=3.9\;\mathrm{T}$. We use the value $J'=0.17\;\mathrm{meV}$ throughout the present paper, which also well reproduces the spectra above 5 T, far enough away from $\mu_{0}H_{\mathrm{c}}=3.9\;\mathrm{T}$, as we shall see below. 

We calculate the uniform longitudinal magnetization $M_{z}$ and staggered transverse magnetization $M_{x}$ for the effective model of Eq.~\eqref{eq:Hamil1Deff} numerically by DMRG in a self-consistent way. The result is shown in Fig.~\ref{fig.4}(b). It is clear that $M_{x}$ behaves as an order parameter for the TAF phase. Furthermore, the transition to the fully-polarized state is calculated to occur around 25 T, which is rather close to the experimentally obtained value of 22.7 T~\cite{Kimura2007}, and which further validates the model. However, we should point out that the existence of an additional phase transition at $\mu_0 H_{1/2}=19.5$~T reported in NMR measurements~\cite{Klanjsek2015} is not observed within the present model Eq.~\eqref{eq:Hamil1Deff}. Hence additional ingredients are likely to be necessary to understand this part of the phase diagram. We will come back to this point in Sec.~\ref{sec:Refinement}.

\section{Results}
\label{sec:Result}

\subsection{Determination of magnetically ordered phases}

Before discussing the spin dynamics, we experimentally check the magnetic ordering stabilized at $T=50$~mK for various fields $\mu_{0} H= 0$, 6, 8, 9.5, and $10\;\mathrm{T}$. The magnetic Bragg peaks are clearly identified in the IN5 neutron data (see Appendix C for the experimental setup), both in $l$-cuts around the $(3, 2, 0)$ position at zero energy [Fig.~\ref{fig.2}(b)] and in the zero energy neutron intensity maps on $(hkl)$ planes with fixed $l$ [Figs.~\ref{fig.2}(c)-\ref{fig.2}(e)]. At $H=0$, nuclear Bragg peaks associated to $h+k+l= (\mathrm{even})$ and magnetic Bragg peaks corresponding to $h+k+l=(\mathrm{odd})$ are visible at $l=0$ in Fig.~\ref{fig.2}(c). This is consistent with the $(1,0,0)$ propagation vector of the LAF phase previously determined. The magnetic Bragg peaks disappear above $T_{N}$ as expected.

At 6 and 8 T, above $H_{\mathrm{c}}$, the magnetic Bragg peaks are no longer visible at $l=0$ and appear instead at the incommensurate positions $l=\pm \delta$ as shown in Fig.~\ref{fig.2}(a). These correspond to a $(1,0,\delta)$ propagation vector with the incommensurate modulation $\delta$, which increases as the field increases. 

Finally, at 9.5 and 10 T, above $H^{*}$, the magnetic signal is back to the $l=0$ slice and associated again to the $(1,0,0)$ propagation vector, which is compatible with the TAF phase. Note that very weak additional $hk0$ peaks are seen with $h$ and $k$ odd, which correspond to forbidden nuclear peaks due to the presence of glide planes perpendicular to ${\bf c}$ in the $I4_{1}/acd$ space group of \bacovo, some of them varying slightly with the field. Their origin is unclear. To conclude, the elastic cuts confirm our previous identification of the successive magnetic ordered phases at very low temperature in \bacovo~ \cite{Canevet2013,Grenier2015prb}, LAF phase up to $\mu_{0} H_{\mathrm{c}} \approx 4\;\mathrm{T}$, followed by the LSDW order up to the spin-flop transition at $\mu_{0} H^{*} \approx 9\;\mathrm{T}$, 
above which the TAF phase is stabilized.

\subsection{Numerical simulations of spin dynamics}

In order to investigate the spin dynamics of \bacovo\ in a longitudinal magnetic field, we numerically calculate the dynamical susceptibility for the effective 1D model of Eq.~\eqref{eq:Hamil1Deff} and confront it to inelastic neutron scattering spectra. The ground state of the system is obtained by DMRG and the retarded correlation function is computed by TEBD. The dynamical susceptibility can be derived as the Fourier transform of this correlation function. The Fourier transform is performed by considering the crystal structure of \bacovo, as done in Ref.~\onlinecite{Takayoshi2018}. The dynamical susceptibility is calculated for each component, specifically the component parallel $S_{\parallel}$ and perpendicular $S_{\perp}$ to the magnetic field. The inelastic neutron scattering spectra $S_{\mathrm{tot}}$ observed experimentally can be represented as a linear combination of $S_{\parallel}$ and $S_{\perp}$ (see Appendix~\ref{sec:numerics} for the details). 

\begin{figure}
\centering
\includegraphics[width=8.5cm]{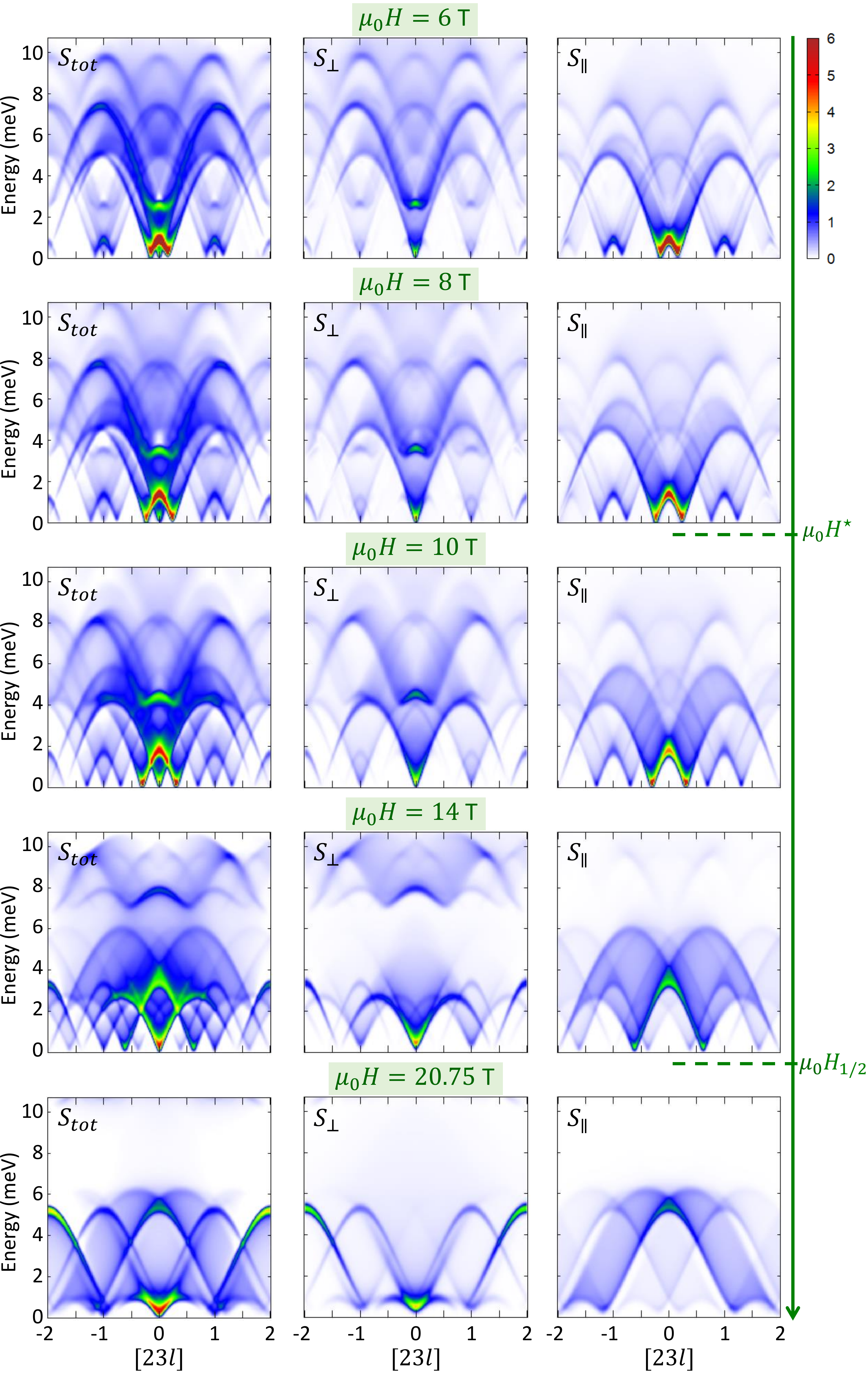}
\caption{Inelastic neutron scattering spectra (dynamical susceptibility) calculated numerically for the model Eq.~\eqref{eq:Hamil1Deff} at various magnetic fields ranging from 6 up to 20.75~T. The total (left), transverse (middle) and longitudinal (right) components are shown. The spin-flop transition $\mu_{0}H^{*}$ detected by neutron diffraction and the transition at half the saturated magnetization $\mu_{0}H_{1/2}$ detected by NMR are indicated by dashed lines.}
\label{fig.5}
\end{figure}

In Fig.~\ref{fig.5}, we show the dynamical spin susceptibility $S_{\parallel}$ and $S_{\perp}$, as well as the total $S_{\mathrm{tot}}$, to be compared with the inelastic neutron scattering spectra. Calculations are performed at various values of magnetic field from 6 T up to 20.75 T, and the slice on the momentum plane $(23l)$ ($-2 \leq l \leq 2$) in the energy regime up to 10.6 meV is shown. The spectra look complex but they are well described by the TLL physics. In the longitudinal component $S_{\parallel}$ of the dynamical susceptibility, the dispersion is expected to become gapless at the wave vectors $q=0$ ($l=\pm 2$) and incommensurate $q=\pi(1\pm m_{z})$ ($l=\pm 2m_{z}$), where $m_{z}\equiv M_{z}/M_{z}^{\mathrm{sat}}$ is the uniform magnetization normalized by the saturated value. 1D spin systems can be transformed to fermionic systems through Jordan-Wigner transformation. In the fermion language, the magnetic field corresponds to a chemical potential and the incommensurate wave number is related to the Fermi wave vector $k_{\mathrm{F}}$. For more details, see~Appendix~\ref{sec:fermionization}. Our calculations agree with this picture with an additional folding into four replica due to the crystal structure of \bacovo\ (which has a four-site periodic spiral form in the ${\bf c}$ direction). In fact, we observe that most of the spectral weight is concentrated in the low energy arch-shaped dispersion that bridges the excitation minima at the incommensurate wave vectors around each reciprocal lattice point. As the external magnetic field is increased, the spectral weight of the longitudinal component decreases continuously and the interval between the gapless points on the arch-shaped dispersion expands. 

The transverse components $S_{+-}$ and $S_{-+}$, whose sum is shown in the $S_{\perp}$ column of Fig.~\ref{fig.5}, also yield a rather complex spectrum. It is seen that $S_{+-}$ is shifted to the high energy side and $S_{-+}$ to the low energy side as the magnetic field is increased~\cite{Chitra1997}. In addition to these spinon continua, multi-magnon excitations are visible at higher energies and attributed to two-string excitations~\cite{Wang2018,Bera2020,Yang2019}. They are pushed to higher energy as the field increases. From the TLL theory, the spinon dispersion appearing in transverse dynamical susceptibility is predicted to become gapless at $q=\pi$ ($l=0$) and $q=\pm \pi m_{z}$ [$l=\pm(2-2m_{z})$]. The spectral weight is concentrated at the commensurate positions corresponding to each reciprocal lattice points of $q=\pi$ in the folded spectrum. It indicates that the alternating correlation is dominant in the transverse component. For two-string excitations, the intensity is strongest around the energy minimum of their dispersion. 

The spin dynamics evolves continuously with increasing the magnetic field up to saturation. No abrupt change of the excitation structure is seen at the critical fields $H^{*}$ and $H_{1/2}$. In addition, despite the existence of long-range magnetic order, the dynamical susceptibility still shows the 1D-like behavior dictated by the TLL theory. 
The spectral weight at low energy is progressively transferred from the longitudinal to the transverse channel, and the latter becomes dominant at high field. Specifically, The low energy intensity at incommensurate positions in $S_{\parallel}$ fades away and is replaced by a strong contribution at the integer $l$ positions in $S_{\perp}$. This result is consistent with the TLL picture, in which the leading order of the correlation function is $\cos[\pi(1+m_{z})r]\,r^{-2K}$ in the longitudinal direction and $\cos(\pi r)\,r^{-\frac{1}{2K}}$ in the transverse direction, where $K$ is the Luttinger parameter and $r$ is the distance (see Appendix~\ref{sec:bosonization}). 

\begin{figure}
\centering
\includegraphics[width=8.5cm]{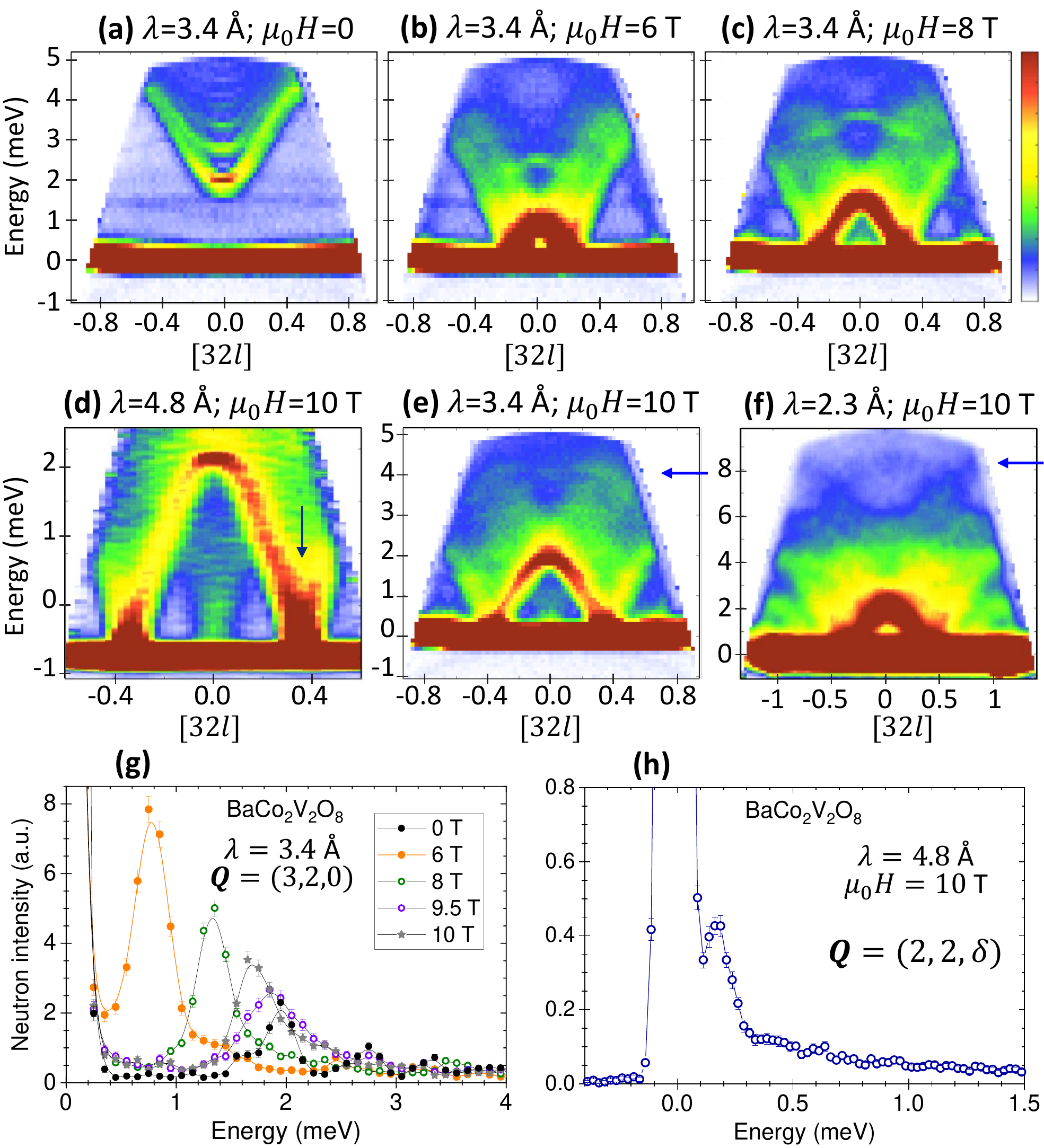}
\caption{Inelastic neutron scattering intensity maps measured on IN5 at 50 mK, showing the dispersion of the excitations around ${\bf Q}=(3,2,0)$ along the ${\bf c}^{*}$ direction, at (a) 0~T, (b) 6~T, (c) 8~T, and (d)-(f) 10~T. The incident wavelength is 3.4~\AA\ for panels (a-c) and (e), 4.8~\AA\ for the panel (d), and 2.3~\AA\ for panel (f), which allows to explore the different energy ranges. The blue arrows indicate the two-string signal in panels (e) and (f). Panel (g) shows energy-cuts at ${\bf Q}=(3,2,0)$ for all measured magnetic fields at 3.4~\AA. Panel (h) shows an energy-cut at ${\bf Q}=(2,2,\delta)$ with $\delta=0.35$, at 10~T and 4.8~\AA\, where the dispersion along the ${\bf c}^{*}$ direction shows a minimum [see the blue arrow on panel (d)]. Note that this gap is slightly larger and thus better visible for the cut $(h,2,\delta)$ with $h=2$ than $h=3$, hence we adopt the value $h=2$ on panel (h).}
\label{fig.3}
\end{figure}

\subsection{Experimental results on spin dynamics}

We then investigate the spin dynamics in the different magnetically ordered phases using inelastic neutron scattering and compare the measured results with the theory (see Appendix~\ref{sec:ExpSetup} for the details of the experimental setup). 

In Figs.~\ref{fig.3}(a)-\ref{fig.3}(f), we show the longitudinal magnetic field evolution of the excitation spectrum measured on IN5 along the ${\bf c}^{*}$ direction on each side of the $(3,2,0)$ position [equivalent to $(2,3,0)$]. Comparison with the results presented in Fig.~\ref{fig.5} shows that the calculations agree quantitatively well with the measurements at 6, 8, and 10~T. In addition, we can identify the properties of excitations by comparing the measured spectra with the numerical results on the dynamical susceptibility of longitudinal and transverse components, which are separately shown in Fig.~\ref{fig.5}. 

At $H=0$ in the LAF phase, a discrete sequence of excitations, with an energy minimum at reciprocal vector (3,2,0), is observed, and is attributed to spinon pairs confined by the interchain interaction~\cite{Grenier2015prl}. Above $H_{\mathrm{c}}$, this spectrum drastically changes. The spectral weight now concentrates in a low energy arch-like excitation that bridges the incommensurate magnetic satellites of the LSDW order at $l=\pm \delta$. This matches very well the expected spin dynamics in the TLL picture. This incommensurability varies with the field as $\delta=2 m_{z}$ with $m_{z}$ the uniform magnetization normalized by the saturated value. The wave vector $q=\pi(1\pm m_z)$ where the dispersion reaches its minimum decrease as the field and $m_{z}$ increases, which indicates that the period of the spatial modulation in the longitudinal spin component becomes longer. The width of this arch-shape of excitation dispersion expands, and the incommensurate positions get further apart from the reciprocal vector position $l=0$ as the field increases. Furthermore, most of the spectral weight comes from longitudinal fluctuations in the LSDW phase, which is the same behavior as the previous measurements up to 6~T~\cite{Faure2019}. Quite remarkably, we do not observe any abrupt change at the flop transition point: the spin excitation spectra change smoothly and continuously through the transition with a further expansion of the arch-like excitation. The energy minima of the dispersion remain at the incommensurate positions predicted by the TLL theory, and thus do not coincide with the TAF Bragg peaks. The continuous increase of the maximum in the arch-like structure at $(3,2,0)$ is shown in Fig.~\ref{fig.3}(g) from 6 to 10~T. 

At 10~T, above the spin-flop transition, the spin excitations were measured with three different incident wavelengths in order to capture features in different energy ranges [Figs.~\ref{fig.3}(d)-(f)]. At 4.8~\AA, the energy resolution is good enough to observe a small energy gap at the incommensurate position $(2,2,\delta)$ ($\delta=0.35$), which is of the order of $\approx 0.2\;\mathrm{meV}$ [Fig.~\ref{fig.3}(h)]. This is similar, within the error bars, to the value reported at 6~T~\cite{Faure2019}. Note that this gap, which is not expected in the perfect TLL picture, stems from the interchain coupling, and has a dispersion in the ${\bf a}^{*}$ (and equivalent ${\bf b}^{*}$) directions. 

Apart from this gap, the above experimental results [Figs.~\ref{fig.3}(a)-(f)] quantitatively agree well with the numerical calculations (Fig.~\ref{fig.5}). This demonstrates that the TLL theory appropriately describes the low energy excitation structure of \bacovo. In Ref.~\cite{Klanjsek2015}, it is reported in the TAF phase that NMR spectra show the incommensurate behavior as well as the AF nature. This feature may be due to the coexistence of the incommensurate dynamics observed in these measurements and described as TLL, and the long-range AF order determined by diffraction. 

At 2.3~\AA, still at 10~T, the spin excitations were measured up to 10~meV. Additional features are observed at the energy 4~meV and 8~meV and at the wave vector $(3,2,0)$. Since these are high-energy excitations, they are beyond the scope of the low energy effective TLL theory but can be captured by numerical calculations as seen in Fig.~\ref{fig.5}. In Ref.~\cite{Bethe1931}, these excitations were described within the algebraic Bethe Ansatz formalism as two-magnon bound states called two-string excitations. The maximum and minimum of their dispersion are observed at the same reciprocal space position due to the folding of the spectra stemming from the four-fold screw chain of \bacovo~\cite{Yang2019}. These features and the whole description of the spectrum in terms of Bethe Ansatz formalism were also reported in the paramagnetic state of SrCo$_2$V$_2$O$_8$~\cite{Bera2020}. We also checked the spin excitations of \bacovo\ in a magnetic field of 10~T at $T=1.1$~K above the N\'eel temperature. No significant change is visible in the excitation structure when comparing the spectra above and below the magnetic ordering temperature. This result confirms that the spin dynamics of \bacovo\ is well described by the TLL theory at low energy, and the behavior is rather unaltered by the long-range order stabilized by the small interchain interaction.

\section{Refinement of the model}
\label{sec:Refinement}

As we have seen above, the model described by Eq.~\eqref{eq:Htot} explains most of the features of \bacovo\ in a longitudinal magnetic field. However, there are still some points to be solved. One of them is an anomalous behavior at the field $\mu_{0} H_{1/2}=19.5\;\mathrm{T}$ in the susceptibility~\cite{Kimura2007} and NMR~\cite{Klanjsek2015} measurements. Another point is that the dispersion of the low-energy excitations, measured by  inelastic neutron scattering, shows an anticrossing. Thus a refinement of the model of Eq.~\eqref{eq:Htot} is necessary, taking into account additional terms in the Hamiltonian. 

In \bacovo, the oxygen octahedra around the Co$^{2+}$ ions along a chain are actually buckled, in a manner that naturally introduces a four-site periodicity (consequence of the four-fold screw axes $4_{1}$ and $4_{3}$ running along the chains). In Ref.~\cite{Kimura2022}, the effects of such a four-site periodic tilting of spin axes due to the crystalline structure were envisaged. This four-site periodic perturbation was thus added to the mean-field model of Eq.~\eqref{eq:Hamil1Deff}, leading to the effective Hamiltonian 
\begin{align}
 \mathcal{H}_{\mathrm{K}}
    =\mathcal{H}_{\mathrm{eff}}
    +\mathcal{H}_{\pi}+\mathcal{H}_{\pi/2},
\label{eq:Hamil_Kimura}
\end{align}
where
\begin{align}
 \mathcal{H}_{\pi}
   =&\,J_{\pi}\sum_{n}(-1)^{n}
   (S_{n}^{+}S_{n+1}^{+}+S_{n}^{-}S_{n+1}^{-})
\nonumber\\
   =&\,2J_{\pi}\sum_{n}(-1)^{n}
   (S_{n}^{x}S_{n+1}^{x}-S_{n}^{y}S_{n+1}^{y})
\\
 \mathcal{H}_{\pi/2}=&J_{\pi/2}\sum_{n}
    \big[\cos(n\pi/2)
   (S_{n}^{x}S_{n+1}^{z}+S_{n}^{z}S_{n+1}^{x})
\nonumber\\
   &\,\qquad\quad +\sin(n\pi/2)
   (S_{n}^{y}S_{n+1}^{z}+S_{n}^{z}S_{n+1}^{y})
   \big].
\label{eq:H_piov2}
\end{align}

\begin{figure}
\centering
\includegraphics[width=7cm]{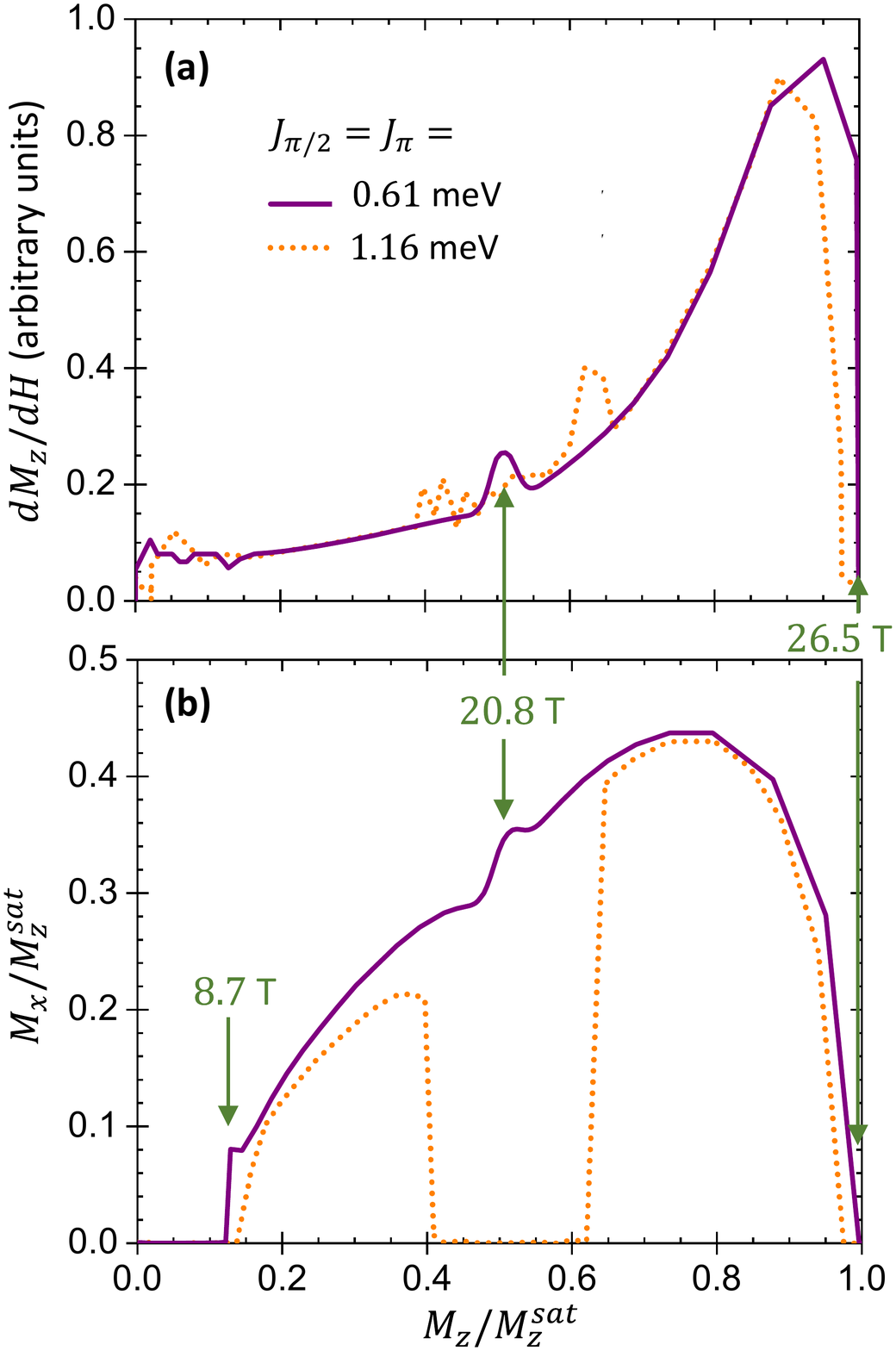}
\caption{(a) Magnetic susceptibility $dM_{z}/dH$ and (b) transverse staggered magnetization $M_{x}$ calculated numerically for the model of Eq.~\eqref{eq:Hamil_Kimura} as a function of $M_{z}$ normalized by its saturated value $M_{z}^{sat}$. The three green arrows locate the corresponding field values at which the various anomalies occur. These values are in good agreement with the experimental ones for $H^{*}$, $H_{1/2}$, and $H_{s}$. The interchain coupling $J'=0.17\; \mathrm{meV}$ and two different kinds of 4-site periodic perturbations, $J_{\pi/2}=J_{\pi}=0.61\; \mathrm{meV}$ (solid purple line) and $J_{\pi/2}=J_{\pi}=1.16\; \mathrm{meV}$ (dotted orange line), are included in the model. The value of $J_{\pi/2}=J_{\pi}=0.61\; \mathrm{meV}$ reproduces well the bump of $dM_{z}/dH$ at half saturation, contrary to the value of $J_{\pi/2}=J_{\pi}=1.16\; \mathrm{meV}$ used in Ref.~\cite{Kimura2022}.}
\label{fig.6}
\end{figure}

\begin{figure}
\centering
\includegraphics[width=8.5cm]{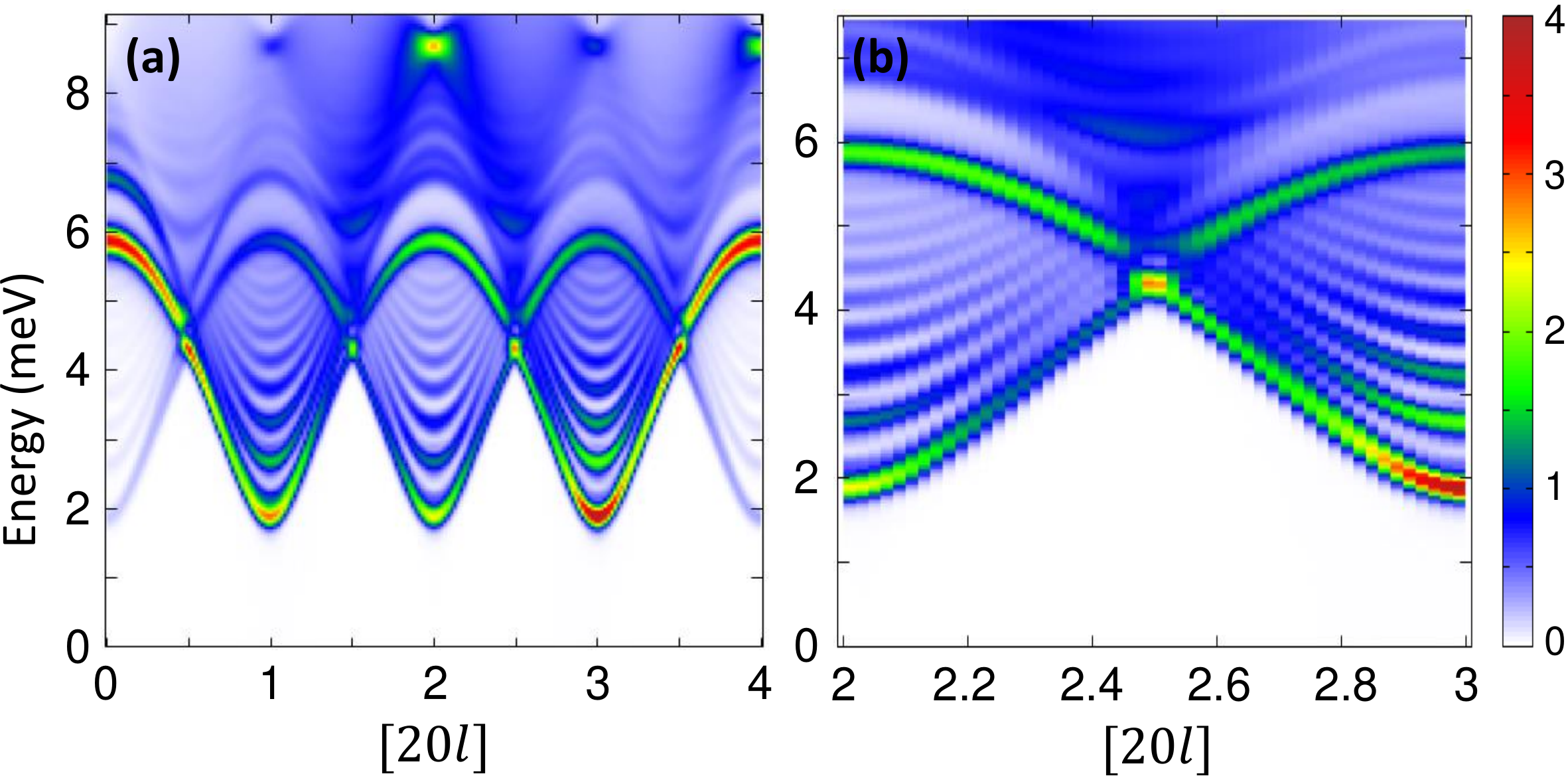}
\caption{(a) Inelastic neutron scattering spectra (dynamical susceptibility) numerically calculated at zero field $H=0$ with an interchain coupling $J'=0.17\; \mathrm{meV}$ and $J_{\pi/2}=J_{\pi}=0.61\; \mathrm{meV}$ for scattering vectors $(2,0,l)$ ($0\leq l\leq 4$). The anticrossing of the dispersion is seen at $l= (\mathrm{integer}+1/2)$. (b) is an expansion of the panel (a) in the $2\leq l\leq 3$ interval.}
\label{fig.7}
\end{figure}

Figure~\ref{fig.6} shows our calculation of the magnetization curve obtained for Eq.~\eqref{eq:Hamil_Kimura}, along with the susceptibility $dM_{z}/dH$ and staggered magnetization $M_{x}$, as a function of $m_z=M_z/M_z^{sat}$. Using $J_{\pi/2}=J_{\pi}=0.61\; \mathrm{meV}$, the susceptibility shows an anomaly at $M_{z}/M_{z}^{\mathrm{sat}}=1/2$, which reproduces the experimental result of Ref.~\cite{Kimura2007}. In addition, we show in Fig.~\ref{fig.7} the calculated dynamical susceptibility, to be compared with inelastic neutron scattering spectra measured along the chain direction in zero field~\cite{Grenier2015prl}. We find that the low energy bands have an anticrossing at the wave vector component $l=($half-odd integer), which corresponds to $q=(2n+1)\pi/4$ ($n$: integer). This indicates that the band folding takes place at $q=n\pi/2$. The four-site periodic perturbation is essential to realize the experimentally observed anticrossing. Actually, the dispersion crosses if there is no such perturbation~\cite{Faure2018}. 

The model of Eq.~\eqref{eq:Hamil_Kimura} with $J_{\pi/2}=J_{\pi}=0.61\; \mathrm{meV}$ can thus \textit{simultaneously} explain the anticrossing of the dispersion observed by inelastic neutron scattering and the anomaly seen in the magnetization curve. Hence, the full model, including this four-site periodic term, with the mean-field treatment of a simple interchain coupling $J'=0.17\; \mathrm{meV}$ fully explains (with the possible exception of the vicinity of 19.5~T, see the discussion below), the observed phase diagram of \bacovo\ as well as fine points observed in the neutron spectra. If additional ingredients are required to be added to this model, they could stem from more complicated interchain couplings as hinted by NMR measurements~\cite{Klanjsek2015}. This point however clearly needs further studies.  

We note that in Ref.~\cite{Kimura2022}, the parameters $J_{\pi/2}=J_{\pi}=1.16$~meV ($J_{\pi/2}/J=J_{\pi}/J=0.2$) were used. With these parameters, however, a new phase characterized by the absence of staggered magnetization ($M_{x}=0$) intrudes into the TAF phase in the chain mean-field theory [Fig. \ref{fig.6}(b)], which is not observed experimentally. We can thus conclude that these parameters seem to be in contradiction with the experimental results. 

\section{Summary and discussion}
\label{sec:Summary}

In this work, we investigated the static and dynamical magnetic properties of the quasi-1D Ising-like compound \bacovo\ under a magnetic field applied along the chains, both from the theoretical and experimental sides. In this compound, thanks to the moderate magnetocrystalline anisotropy, quantum phase transitions are achieved at accessible magnetic fields. The magnetic field leads to a very rich phase diagram and physical properties. 

Our combination of analytical and numerical techniques allows to determine a model which reproduces the observed experimental features with the possible exception of those observed at 19.5~T. In particular, we reproduce the sequence of phases seen in Fig.~\ref{fig.1} up to 19.5~T and the observed neutron spectra up to 10~T. Starting from a longitudinal antiferromagnet at zero magnetic field, as the longitudinal field is increased, the anisotropy gap closes at about 4 T, and the incommensurate LSDW phase and TAF phase successively appear. Although the LSDW and TAF phases are magnetically long-range ordered, the dynamics of the system can be described by the TLL physics. In particular, the inversion of the dominant spin-spin correlation between the longitudinal incommensurate component and the transverse staggered one could be investigated. This inversion indicates that the correlation length becomes larger in the transverse direction than in the longitudinal direction above the spin-flop transition, which has a straightforward consequence on the static properties. With the weak interchain interaction, a long-range order associated to the longer correlation length is stabilized, i.e., longitudinal and incommensurate below $H^{*}$, transverse and staggered above. A large part of the magnetic moment is still fluctuating though because of the 1D nature of the system. The stabilized long-range order is associated with excitations on shorter range length scales of the spin fluctuations as already observed in frustrated magnets~\cite{Chitra1995}. Therefore, the spin dynamics sustains both transverse and longitudinal fluctuations, and the magnetic order is therefore expected to reflect the dominant one at low energy~\cite{Klanjsek2015}, which changes from longitudinal one at low field, as observed in the LSDW phase~\cite{Faure2019} to transverse one at high field in the TAF phase. This results in drastically different magnetic 3D long-range orderings in contrast to very similar 1D spin dynamics close to $H^{*}$. Such an apparent disconnection of the static and dynamical properties is another nice property of the rich physics observed in the low-dimensional quantum antiferromagnet \bacovo, which is proved to be a precious material to test universal behaviors described by the TLL theory. 

However, \bacovo\ exhibits also deviations from the above-mentioned generic behavior in some aspects. We thus refined the model to describe properly these observed deviations by adding four-site periodic terms related to the tilting of the octahedra along the chain axis. This allowed to reproduce the anticrossing observed in the neutron spectra and the anomaly in susceptibility around $\mu_{0}H_{1/2}=19.5\;\mathrm{T}$. Despite this success of the present theoretical approach, some aspects of the phenomena observed around $H_{1/2}$ remain to be understood. The data from NMR advocate for an additional transition at this field [Fig.~\ref{fig.1}(a)], inferred from the peak of the NMR relaxation rate $1/T_{1}$~\cite{Klanjsek2015}. This transition is tentatively attributed to a strong variation of the interchain coupling due to the incommensurability and to their sign change between the TAF and "?" phases. The question of the interchain coupling is at that stage 
largely open and is clearly an important issue for future studies. A strong variation of the couplings with incommensurability would a priori be surprising in view of the fact that we can quite quantitatively describe the TAF neutron spectra with a single and constant effective interchain coupling between 5~T and 19~T in our mean-field approach. Hence, the dip of 3D ordering temperature seen at 19~T in the NMR and the question of the existence of two separate phases remains open. One way of investigation to be done in future studies is to determine if the more complete model we have introduced could explain the dip of 3D ordering temperature.  Additional experimental data around 19~T would be needed, as well as a further knowledge of the interchain coupling.

\acknowledgements
We thank J. Debray, A. Hadj-Azzem, and J. Balay for their contributions to the crystal growth, cut, and orientation. We acknowledge ILL for allocation of neutron beamtime. S.T. is supported by JSPS KAKENHI Grant No. JP21K03412 and JST CREST Grant No. JPMJCR19T3, Japan. This work was supported in part
by the Swiss National Science Foundation under Division II.

\appendix

\section{Jordan-Wigner fermionization and correspondence to \bacovo}

\label{sec:fermionization}

\begin{figure}[tb]
\begin{center}
\includegraphics[width=\linewidth]{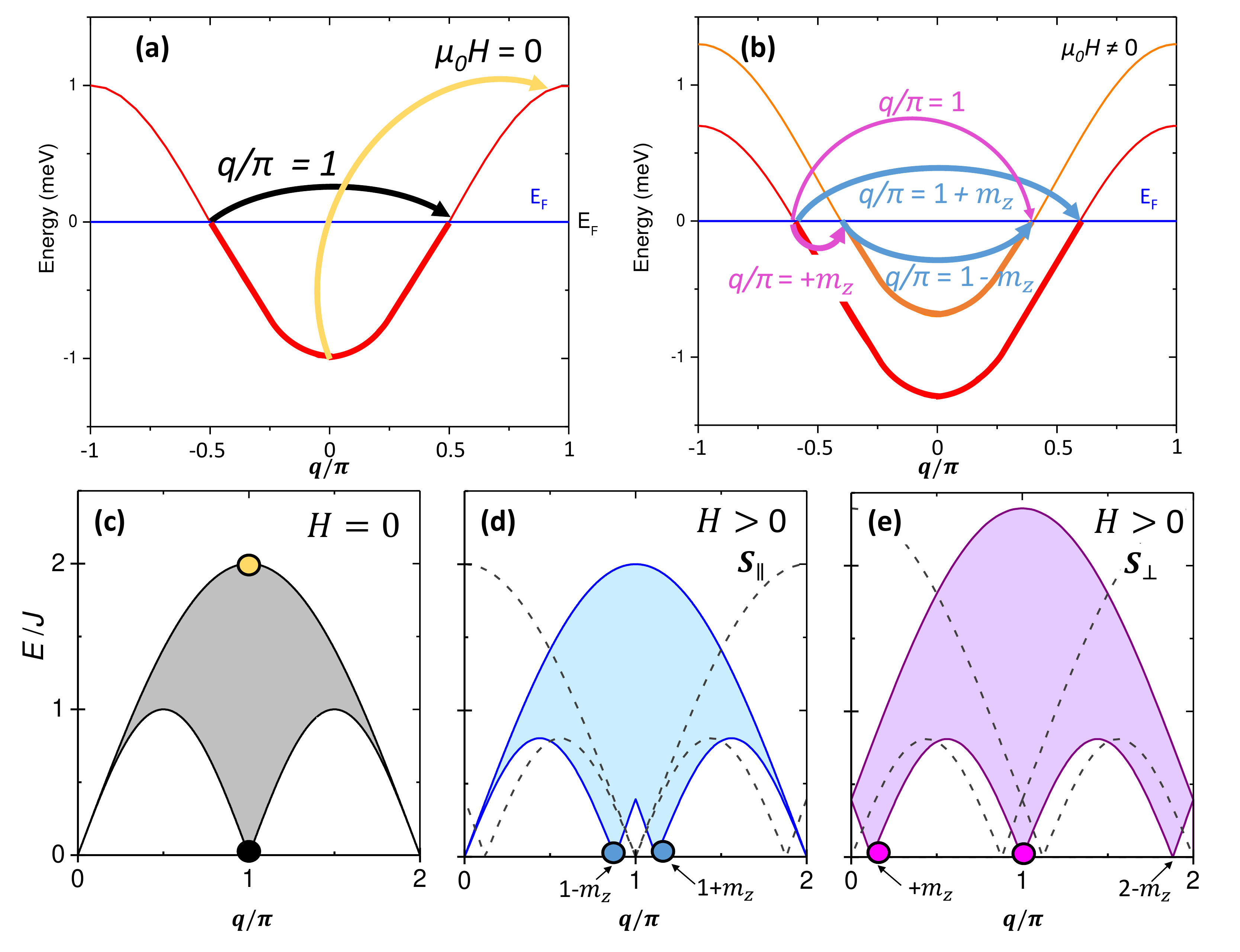}
\caption{(a) Particle-hole band at zero field obtained through the Jordan-Wigner transformation. Two specific particle-hole excitations are pointed out by the black and yellow arrows, which correspond to zero-energy and maximum energy excitation, respectively. The Fermi level is denoted by $E_{\mathrm{F}}$, which is set to zero. The fermions are located below $E_{\mathrm{F}}$ (thick line) while the holes are located above (thin line). (b) The applied longitudinal field $H$ is equivalent to a chemical potential which splits the degeneracy of the electron-hole bands. The intraband and interband zero-energy excitations, which respectively corresponds to longitudinal $S_{\parallel}$ and transverse $S_{\perp}$ fluctuations, are pointed out by the blue and pink arrows. The incommensurability arises from the splitting of the bands. (c) Two-spinon continuum expected for the spin-1/2 XY chain at zero-field. Black and yellow points are the excitations corresponding to the black and yellows arrows. (d),(e) Dynamical susceptibility spectrum of the longitudinal ($S_{\parallel}$) and transverse ($S_{\perp}$) fluctuations expected for an spin-1/2 XY chain in the case of $H > 0$. We point out some of the incommensurate positions where the gap closes by the blue and pink points, which are associated to the blue and pink arrows. The evolution of excitation spectra shown in panels (c)-(e) looks similar even in the presence of nearest-neighbor interaction, in particular for Ising anisotropic system in the TLL region $H_{\mathrm{c}} < H < H_{\mathrm{sat}}$, but the energy, which depend on the anisotropy, is then renormalized, e.g., multiplied by $\pi/2$ for the Heisenberg case. The black dashed lines in (d) and (e) show the effect of folding from the crystal structure of \bacovo.}
\label{Fig8}
\end{center}
\end{figure}

The incommensurability appearing in the spin-dynamics for an AF spin-1/2 chain under a longitudinal magnetic field can be understood in terms of spinless fermions through the Jordan-Wigner transformation: 
\begin{align}
\begin{split}
 S_{j}^{+} =& c_{j}^{\dagger}\prod_{i=1}^{j-1} (1-2n_{i}), \quad
 S_{j}^{-} = c_{j}\prod_{i=1}^{j-1} (1-2n_{i}),\\
 S_{j}^{z} =& n_{j}-\frac{1}{2},
\end{split}
\end{align}
where $c_{j}$ ($c_{j}^{\dagger}$) is the fermion annihilation (creation) operator and $n_{j}=c_{j}^{\dagger}c_{j}$ is the number operator. The spin $S^{z}=1/2$ ($S^{z}=-1/2$) corresponds to an electron (a hole). For simplicity, we consider the XY spin chain here, which transforms into a free electron-hole band theory. We apply this fermionization for a spin-1/2 chain with a constant lattice spacing $a$ ($=c/4$ for \bacovo) under a field along the $z$ axis. 
Firstly, we change from the antiferromagnet to ferromagnet through the spin rotation  
$S_{j}^{x,y} \to (-)^{j}S_{j}^{x,y}$, 
and the Hamiltonian becomes 
\begin{align}
 \mathcal{H} &= -\frac{J}{2} \sum_{j}
   \big(S_{j}^{+}S_{j+1}^{-}+S_{j}^{-}S_{j+1}^{+}\big)
   -\mu_{\mathrm{B}} g_{zz} H \sum_{j} S_{j}^{z}
\end{align}
Then through the Jordan-Wigner transformation, we derive 
\begin{align}
 \mathcal{H} = -\frac{J}{2} \sum_{j}
   \big(c_{j}^{\dagger} c_{j+1} + c_{j+1}^{\dagger}c_{j} \big)
   -\mu_{\mathrm{B}} g_{zz} H \sum_{j} n_{j},
\label{eq:JordanWignerField}
\end{align}
where the constant is omitted. After the Fourier transform, we obtain 
\begin{align}
\begin{split}
 &\mathcal{H} = \sum_{k} E(k) c_{k}^{\dagger} c_{k} \\ 
 &E(k) = -J \cos(ka) - \mu_{\mathrm{B}} g_{zz} H 
\end{split} 
\end{align}
At zero field, the fermion band is half-filled, as depicted in Fig.~\ref{Fig8}(a). The continuum of spinons is understood is terms of the density of states of a fermion and a hole, as shown in Fig.~\ref{Fig8}(c).

Let us now turn to the case where a magnetic field is applied. The magnetic field lifts the degeneracy of the spin states according to the value of $\sum_{j}S_{j}^{z}$ by Zeeman splitting. In the fermionic language of Eq.~\eqref{eq:JordanWignerField}, the longitudinal magnetic field is recast to a chemical potential, which modifies the filling of the band. Instead of considering one spinless fermion band and shifting the chemical potential, the situation can be understood as two bands, one for particles and one for holes, shifting with respect to the Fermi level in opposite directions, as represented in Fig.~\ref{Fig8}(b). 

The longitudinal excitations $S_{\parallel}$ conserving the total number of particles, i.e., intraband excitations (without a change of magnetization from the ground state: $\Delta S^{z} = 0$) give rise to incommensurate fluctuations reaching zero energy at $q = \pi(1 \pm m_{z})$. $m_{z}= 1 - (2/\pi)\arccos (\mu_{\mathrm{B}} g_{zz} H / J)$ is the magnetization normalized by its saturation value. In addition commensurate fluctuations appear at $q = 0$ and $2 \pi$ [see Fig.~\ref{Fig8}(b)]. The transverse excitations $S_{\perp}$ changing the number of particles, i.e. interband excitations (leading to a change of magnetization: $\Delta S^{z} = \pm 1$) give rise to fluctuations reaching zero energy at incommensurate wave numbers $q=\pi m_{z}$ and $\pi (2-m_{z})$ in addition to $q = \pi$ (see Fig.~\ref{Fig8}(b) and also Ref.~\cite{Chitra1997}). From all possible intraband and interband excitations, as in the zero-field case, one can reconstruct the longitudinal $S_{\parallel}$ and transverse $S_{\perp}$ dispersion spectra expected for a spin-1/2 chain in a magnetic field along the $z$ axis [see Figs.~\ref{Fig8}(d) and \ref{Fig8}(e)].

The crystalline unit cell of \bacovo\ encompasses four chains, and each chain winds in a spiral around the $c$ axis with a period of four spins. As a result, the correspondence between $l$ and $q$ is $l=2q/\pi$. The incommensurability in $S_{\parallel}$ at $q/\pi=(1\pm m_{z})$ appears at $l=4n-2\pm\delta$ ($n$ integer) with $\delta=2m_z$. In the same way, the incommensurability in $S_{\perp}$ at $q/\pi=\pm m_{z}$ appears at $l=4n\pm\delta$. 
However, the situation is more complicated due to the details of the structure of \bacovo. The actual positions of the \co\ atoms light on replicas of the main spectrum described above, shifted by $\Delta l=0,1,2,$ and $3$. Both $S_{\parallel}$ and $S_{\perp}$ become thus visible around each $l$ integer, as can be also observed in the calculations shown in Figure~\ref{fig.5}.
The spectral weight also depends on the values of $h$ and $k$ due to the interchain interaction. The actual magnetic structure gives rise to Bragg peaks at $(hkl)$ positions such that $h+k+l=2n+1$. As a result, the spectrum is best observed at $(320)$ or equivalently $(230)$. In this case, since the antiferromagnetic wave number $q=\pi$ corresponds to $l=4n$, incommensurability in $S_{\parallel}$ ($S_{\perp}$) is observed at $l=4n\pm\delta$ ($l=4n-2\pm\delta$).

\section{Bosonized field theory}
\label{sec:bosonization}

The XXZ model 
\begin{align}
 \mathcal{H}_{\mathrm{XXZ}}
   =&J\sum_{j}[\epsilon (S_{j}^{x}S_{j+1}^{x}
   +S_{j}^{y}S_{j+1}^{y})
   +S_{j}^{z}S_{j+1}^{z}]
\label{eq:Hamil_XXZ_App}
\end{align}
can be analyzed through the mapping to an effective field theory by bosonization~\cite{giamarchi2004}. The spin operators are represented by bosonic scalar fields as
\begin{align}
 S_{j}^{z}&=-\frac{c/4}{\pi}\partial_{z}\phi(z)
   +a_{1}(-1)^{j}\cos(2\phi(z))+\cdots,\\
 S_{j}^{+}&=e^{-i\theta(z)}
   [b_{0}(-1)^{j}+b_{1}\cos(2\phi(z))+\cdots],
\end{align}
where $z=j(c/4)$ is the coordinate along the crystal ${\bf c}$ axis with the lattice constant $c$, and $a_{0}$, $b_{0}$, and $b_{1}$ are nonuniversal coefficients. Using the dual bosonic fields $\phi(z)$ and $\theta(z)$, we can recast the Hamiltonian Eq.~\eqref{eq:Hamil_XXZ_App} into
\begin{align}
 \mathcal{H}_{\mathrm{bos}}&=\frac{v}{2\pi}\int dz
   \Big[\frac{1}{K}(\partial_{z}\phi(z))^{2}
   +K(\partial_{z}\theta(z))^{2}\Big]\nonumber\\
   &\quad-\lambda\int dz\cos(4\phi(z))
   +\cdots,\nonumber
\end{align}
where $v$ is the spinon velocity, $K$ the Luttinger parameter, and $\lambda$ a nonuniversal constant. Since the scaling dimension of the $\cos(4\phi(z))$ term is $4K$, this term is relevant in the easy-axis ($\epsilon<1$, $K<1/2$) regime and works as a potential to pin the field $\phi(z)$. Hence the system has a spin excitation gap. When the longitudinal magnetic field is applied, the Hamiltonian becomes
\begin{align}
 \mathcal{H}_{\mathrm{XXZ}}-h\sum_{j}S_{j}^{z}.
\end{align}
Thus, the bosonized Hamiltonian is 
\begin{align}
 \mathcal{H}_{\mathrm{bos}}
    +\frac{h}{\pi}\int dz \partial_{z} \phi(z)
\end{align}
While the $\cos(4\phi(z))$ term pins the field $\phi(z)$, the $\partial_{z} \phi(z)$ term tries to shift it, thus the competition happens. When $h$ is increased and the spin gap closes, the $\partial_{z} \phi(z)$ term becomes dominant, the magnetization $\braket{S_{j}^{z}}=-\braket{\partial_{z} \phi(z)}/\pi$ becomes nonzero, and the Pokrovsky-Talapov commensurate-incommensurate phase transition takes place~\cite{Pokrovsky1979,Chitra1997,Giamarchi1999}. When the magnetization grows, the system is gapless and can be described as a TLL. 

The longitudinal and transverse spin-spin correlation functions in the magnetized system are written as
\begin{align}
 \braket{S_{j}^{z}S_{0}^{z}}
   =&\braket{S_{j}^{z}}\braket{S_{0}^{z}}+\frac{K}{2\pi^{2}}\,z^{-2}
\nonumber\\
     &+C_{1}\cos(\pi(1+m_{z})z)\,z^{-2K}
      +\cdots\\
 \braket{S_{+}^{z}S_{-}^{z}}
   =&C_{2}\cos(\pi z)\,z^{-\frac{1}{2K}}
\nonumber\\
     &+C_{3}\cos(\pi m_{z}z)\,z^{-2K-\frac{1}{2K}}
     +\cdots,
\end{align}
where $m_{z}=\braket{S_{j}^{z}}/M_{z}^{\mathrm{sat}}$ is the magnetization normalized by its saturated value. While the magnetization increases from 0 to the saturation value, the Luttinger parameter $K$ increases from $1/4$ to $1$. In the regime of $1/4<K<1/2$ ($1/2<K<1$), the longitudinal (transverse) correlation is dominant and the LSDW (TAF) phase is stabilized when the interchain interaction is introduced.

\section{Details of the numerical calculations}
\label{sec:numerics}
In this appendix, we explain the details of the numerical methods~\cite{Faure2018,Takayoshi2018}. We treat the interchain interaction in terms of the mean-field approximation and we derive an effective 1D Hamiltonian 
\begin{align}
 \mathcal{H}_{\mathrm{eff}}
   =& J \sum_{n} [\epsilon
 ( S_{n}^{x} S_{n+1}^{x} 
  + S_{n}^{y} S_{n+1}^{y})
  + S_{n}^{z} S_{n+1}^{z}]
\nonumber\\
   &- g_{zz}\mu_{B}\mu_{0} H\sum_{n} S_{n}^{z}
     + J'\sum_{n}
     \braket{{\bf S}_{n}}\cdot
     {\bf S}_{n},
\label{eq:Hamil1Deff_App}
\end{align}
where $\braket{{\bf S}_{n}}$ is calculated iteratively and determined self-consistently. We set the parameters $J=5.8$~meV, $\epsilon=0.53$ and 
$g_{zz}=6.07$~\cite{Faure2019}. 

We calculate the dynamical susceptibility numerically. First, the ground state is obtained by DMRG~\cite{white1992}, then its time-evolution is calculated by TEBD~\cite{vidal2003}. In this way, we can evaluate the spin-spin retarded correlation function 
\begin{align}
 C_{\mathrm{R}}^{\alpha\beta}({\bf r},t)
   =-i\vartheta_{\mathrm{step}}(t)
     \braket{[S^{\alpha}({\bf r},t),S^{\beta}({\bf 0},0)]}
\end{align}
for the Hamiltonian Eq.~\eqref{eq:Hamil1Deff}, where $\vartheta_{\mathrm{step}}(t)$ is the step function. We take as the system's size $N=200$, the time interval $0\leq t\leq 60J^{-1}$, the time-discretization $dt=0.05J^{-1}$, and the bond dimension of matrix product states $M=60$. The dynamical susceptibility is obtained from the Fourier transform of the retarded correlation function as 
\begin{align}
 \chi_{\alpha\beta}({\bf Q},\omega)
   =-\mathrm{Im}\int dt\sum_{{\bf r}}
   e^{i(\omega t-{\bf Q}\cdot{\bf r})}
   C_{\mathrm{R}}^{\alpha\beta}({\bf r},t).
\label{eq:DSF}
\end{align} 
For the Fourier transform in Eq.~\eqref{eq:DSF}, the summation is taken over the actual positions ${\bf r}$ of Co$^{2+}$ ions. 

The differential neutron scattering cross section $S({\bf Q},\omega)$ can be related to the dynamical susceptibility $\chi_{\alpha\beta}({\bf Q},\omega)$ as 
\begin{align}
 &S_{\mathrm{tot}}({\bf Q},\omega)
 \nonumber\\
 &=\frac{|{\bf q}'|}{|{\bf q}|}\sum_{\alpha,\beta=x,y,z}
   \Big(\delta_{\alpha\beta}-\frac{Q_{\alpha}Q_{\beta}}{|{\bf Q}|^{2}}\Big)
   |f(\textbf{Q})|^{2}\chi_{\alpha\beta}({\bf Q},\omega),
\label{eq:CrossSec}
\end{align} 
where $f({\bf Q})$ is the magnetic form factor and ${\bf q},{\bf q}'$ are the incident and scattered wave vectors, respectively, and ${\bf Q}={\bf q}-{\bf q}'$. To see the longitudinal and transverse excitations independently, we also define 
$S_{\parallel}({\bf Q},\omega)=\chi_{zz}({\bf Q},\omega)$ 
and 
$S_{\perp}({\bf Q},\omega)=\chi_{xx}({\bf Q},\omega)$. 

To reproduce the susceptibility anomaly at $M_{z}/M_{z}^{\mathrm{sat}}=1/2$ and the band anticrossing at the wave number $l=(\mathrm{half \; integer})$, the four-site periodic perturbation given in Ref.~\cite{Kimura2022} is taken into account. Thus we consider the Hamiltonian 
\begin{align}
\mathcal{H}_{\mathrm{K}}
    =\mathcal{H}_{\mathrm{eff}}
   +\mathcal{H}_{\pi}+\mathcal{H}_{\pi/2},
\end{align}
where
\begin{align}
 \mathcal{H}_{\pi}
   &=\,J_{\pi}\sum_{n}(-1)^{n}
   (S_{n}^{+}S_{n+1}^{+}+S_{n}^{-}S_{n+1}^{-})
\nonumber\\
   &=\,2J_{\pi}\sum_{n}(-1)^{n}
   (S_{n}^{x}S_{n+1}^{x}-S_{n}^{y}S_{n+1}^{y})
\end{align}
and 
\begin{align}
 \mathcal{H}_{\pi/2}=J_{\pi/2}&\sum_{n}
    \big[\cos(n\pi/2)
   (S_{n}^{x}S_{n+1}^{z}+S_{n}^{z}S_{n+1}^{x})
\nonumber\\
   &+\sin(n\pi/2)
   (S_{n}^{y}S_{n+1}^{z}+S_{n}^{z}S_{n+1}^{y})
   \big].
\end{align}
The procedure to obtain the dynamical spin susceptibility is the same as in the case of $\mathcal{H}_{\mathrm{eff}}$. 

\section{Experimental setup}
\label{sec:ExpSetup}

The single-crystal of \bacovo\ was synthesized by the floating zone method in an image furnace \cite{Lejay2011}.

Time-of-flight experiments were performed on the IN5 spectrometer~\cite{Ollivier2011} at the Institut Laue Langevin in a 10~T vertical magnet with a dilution insert allowing to cool down to 50 mK~\cite{GrenierILLdata}. The crystal was oriented with the ${\bf c}$ axis vertical and the horizontal scattering plane (${\bf a}^{*}$, ${\bf b}^{*}$). The large vertical covering of the detector ($\pm 20^{\circ}$) allowed to probe a large portion of the reciprocal space including in the ${\bf c}^{*}$ direction. The elastic and inelastic signals could be probed in the same experiment. Three different wavelengths, 2.3 \AA, 3.4 \AA, and 4.8 \AA\ were used to access different energy ranges.

\end{document}